\documentclass[preprint,amssymb,showpacs,supersriptaddress,floatfix]{revtex4-2}
\usepackage{graphicx}
\usepackage{dcolumn}
\usepackage{bm}
\usepackage[utf8]{inputenc}
\usepackage[T1]{fontenc}
\usepackage{mathptmx}
\usepackage{etoolbox}
\usepackage{float}
\usepackage{amsmath,amssymb}
\usepackage[dvipsnames]{xcolor}
\usepackage{subcaption}
\makeatletter
\def\@email#1#2{
 \endgroup
 \patchcmd{\titleblock@produce}
  {\frontmatter@RRAPformat}
  {\frontmatter@RRAPformat{\produce@RRAP{*#1\href{mailto:#2}{#2}}}\frontmatter@RRAPformat}
  {}{}
}
\makeatother

\usepackage{color}

\usepackage[utf8]{inputenc}
\usepackage[T1]{fontenc}

\begin{document}



\title{On the origin of filamentary resistive switching in oxides-based memristive devices}
\author{R. Leal Martir$^{1,2}$}
\author{E. A. Jagla$^{2}$}
\author{D. Rubi$^{3}$}
\author{M. J. Sánchez $^{1,2,*}$}

\affiliation{$^{1}$Instituto de Nanociencia y Nanotecnología (INN),CONICET-CNEA, nodo Bariloche, 8400 San Carlos de Bariloche, Río Negro, Argentina.\\
$^{2}$Centro Atómico Bariloche, Instituto Balseiro (UNCuyo), CONICET, 8400 San Carlos de Bariloche, Río Negro, Argentina.\\
$^{3}$Instituto de Nanociencia y Nanotecnología (INN), CONICET-CNEA, nodo Buenos Aires, Argentina.\\}

\email{maria.sanchez@ib.edu.ar}

\begin{abstract}

The control and manipulation of filamentary resistive switching (FRS)  is essential for practical applications  in fields like non-volatile memories and neuromorphic computing. However, key aspects of the dynamics of conductive filament formation and their influence on device resistance remain incompletely understood. In this work we study  FRS in binary oxides based memristors by  investigating the dynamics of oxygen vacancies (OV) on a two dimensional lattice and their role in forming low-resistance paths that facilitate the transition between high and low global resistance states. We reveal that the mere formation of an OV percolation path is insufficient to induce a transition to a low-resistance state. Instead, an OV concentration exceeding a critical threshold across all sites in the filament is required to generate a low-resistivity conducting path.
Furthermore, we simulate the impact of static defects -which block OV migration and would correspond to voids in real porous samples-, on filament formation. We show that there is a range of defect density values where OV percolate through the sample, leading to the formation of OV filaments, but conductive paths  remain absent. Additionally, a small concentration of defects can reduce the final value of the low-resistance state, thereby increasing the ON-OFF ratio.
These findings provide valuable insights into optimizing defective nanomaterials with memristive properties, which are crucial for advancing in-memory and neuromorphic computing technologies.

\end{abstract}

\maketitle

 \section{INTRODUCTION}

Memristive devices are expected to be key components in the development of neuromorphic hardware \cite{yu_2017,pre_2015}, intended to outperform  current software-based algorithms by mimicking the structure and information processing mechanism of the brain \cite{zhang_2020, kendall_2020}.
Typical devices are metal/insulator/metal (MIM) structures  of micro-nanometer size  that exhibit the resistive switching (RS) effect - i.e. the reversible change of  the resistance upon the application of an electrical stimulus \cite{saw_2008, iel_2016}. 

Since almost two decades, numerous studies  of RS in devices where the insulator is a transition metal oxide (TMO) -mainly binary compounds such as TiO$_2$, NiO, HfO$_2$, VO$_2$ and Ta$_2$O$_5$, among others\cite{nandi_2020, ghe_2013, Fer_2020, son_2008, alvarez_2024, wang_2015_2, sharoni_2008}- have been reported as bipolar and  atributted to the formation  and retraction/rupture of nanoscale OV conducting   filaments  upon polarity reversal of the applied stimulus \cite{yan_2008, zhu_2024, kim_2020, park_2015, chen_2015}.
 In binary TMO, it is well known that the local resistivity decreases with the OV content \cite{bao_2023, lu_2012, arif_2017},
and thus  the formation/disruption of   OV  filaments seems to be at the origin of  the resistive  changes \cite{saw_2008, waser_2010, bae_2012, balatti_2013}.

Although  in situ detection and visualization of the precise location of conducting filaments are still challenging, filamentary RS has been experimentally characterized through various methodologies, including surface measurement techniques like conductive atomic force microscopy (c-AFM) and thermal mapping \cite{son_2008, nandi_2020, celano_2013, isaev_2023}, which reveal localized conduction paths. Additional insights have been gained through other imaging techniques such as scanning electron microscopy (SEM), and scanning transmission electron microscopy (STEM), which have provided direct evidence of conductive filaments  \cite{kwo_2010,yang_2012_2, peng_2012, krishnan_2016,cheng_2021}. Indirect metrics, such as the absence of scaling between resistance levels and device area, further corroborate the presence of filamentary structures \cite{sas_2016}.

In general, theoretical works on filamentary RS primarily focus on filament formation, either overlooking the underlying physical mechanisms \cite{brivio_2017, xing_2016, arijit_2020,maldonado_2024} or presenting an incomplete picture of  how these mechanisms impact the device's overall resistance \cite{xu_2023}.
On the other hand, certain models such as the Voltage Enhanced OV migration (VEOV) model \cite{roz_2010} have proven effective in reproducing  the  RS effect, by correlating the dynamics of OV  with the different resistance states \cite{ghe_2013, Fer_2020, LMartir_2023}. However, its  one dimensional (1d) nature has precluded a thorough exploration of the role of localized defects  that might be present in the sample which, as we will show, have a key impact on the formation and subsequent dynamics of  filaments.

In order to overcome this limitation, the present study employs a two-dimensional (2d) resistor network  model, in which  the resistivities of the network   depend on  the OV concentration. The model, shortnamed  OVRN (for Oxygen Vacancies Resistive Network), allows us to examine OV dynamics in a spatially resolved framework, providing new insights into how initial OV distributions, and it subsequent dynamics upon an external applied stimulus, determine the  filament formation. In this way  the OVRN captures the  changes in the resistivities of  the network, and the concomitant evolution of the resistance of the device.  
As we will show along the present work, inhomogeneities in the initial OV distribution are at the origin of conducting filament formation. In addition   the duration of the applied electrical stimulus  will be critical,  affecting the  filament size,  width and stability.
The OVRN model also opens the possibility to explore different types of localized defects  such as grain boundaries \cite{bejtka_2020, clarke_2016, yan_2016} or porosity \cite{chakrabarti_2021}, and their role relevant to the RS effect- in particular  the influence of such defects in  the filamentary RS. 
A recent work by Xu et al. \cite{xu_2023} studied the filament growth processes in electrochemical metallization cells  employing kinetic Monte Carlo 
simulations, identifying three different growth modes and concluding 
that the density of high mobility regions is a relevant parameter to distinguish these modes.
Our OVRN model permits to define regions in the sample  in which the
OV concentration is maintained at zero, therefore producing an  effect equivalent to that studied in \cite{xu_2023}. 
However, we will  make a key distinction between an OV filament (OVF) and a low resistivity conductive path (CP). 
With this distinction at hand we define two  thresholds: one for  OV percolation and the other for CP formation, which are identified and analyzed for different densities of localized defects. 
Furthermore, an interesting outcome of our study is a scaling effect in the temporal response of the resistance  for different applied stimuli, that breaks down when the concentration of static defects exceeds a threshold.




\section{The OVRN model: Oxygen vacancies dynamics on a  2d resistor network}

We start by introducing the OVRN  model, which  describes the resistivity change  of a resistor network (RN) in terms of the migration of OV (positively charged) under an external electrical stimulus, making use of the   well known  dependence of the oxide's resistivity with the OV content. In this way, following the dynamics of  OV profiles  for an applied stimulus, $V$, we compute  the changes in the resistance of the sample as a function of time.  

\begin{figure}[H] 
\centering
\includegraphics[width=1\linewidth]{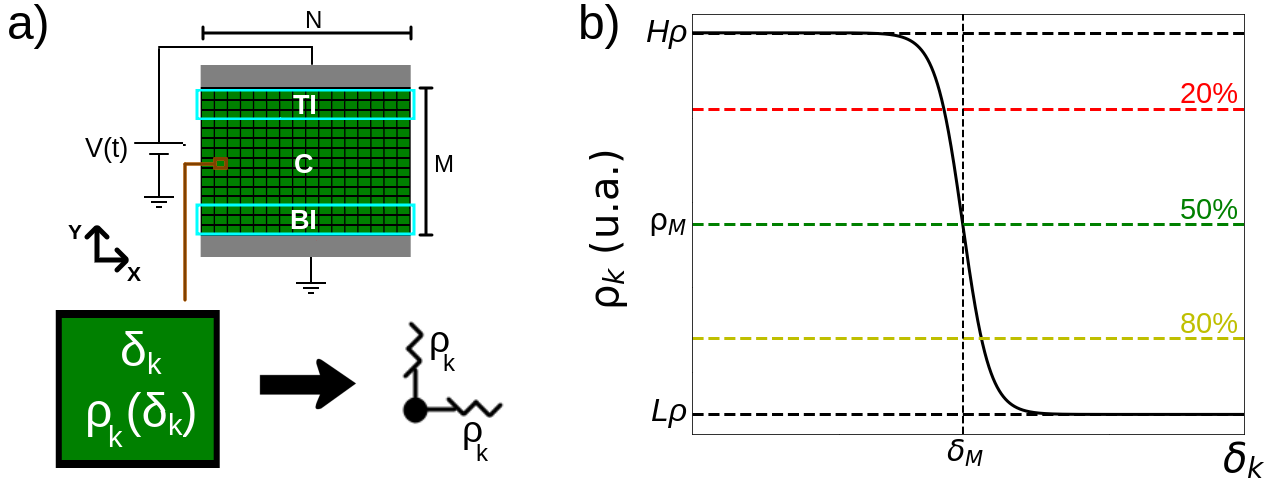}
\caption{ (a) Sketch of the  array of  NxM (N=60, M=30) nanodomains considered in the OVRN model. The TI, C and BI regions are highlighted. Each nanodomain $k$  is identified with a site in the RN, with  associated  horizontal and vertical resistivities (both equal),  $\rho_k$. (b) Resistivity of a nanodomain ($k$) as a function of its OV concentration $\delta_k$. Three different resistivity  threshold   are shown and indicated by the coloured dashed lines. See text for more details.}
\label{Fig1}
\end{figure}

The OVRN considers the sample  as a  rectangular array of $N \times M$ nanodomains in the $x-y$ plane, each one labelled by an index $k$, $1\le k\le N \times M$.
The resistivity of each nanodomain $\rho_{k}$, is  a function  of the  local OV concentration $\delta_{k}$ (i.e., the  number of OV in domain $k$ divided by the total number of OV in the sample). This function is taken to be of the form
\begin{equation}\label{eqnre}
    \rho_{k} = \rho_M -\rho_m \tanh[A(\delta_{k} - \delta_{M})] ,
\end{equation}
where $ \rho_m \equiv (H\rho-L\rho)/2$ and $ \rho_M \equiv (H\rho+ L\rho)/2$,  are defined in  terms of the  minimum, $L\rho$, and maximum, $H\rho$, resistivities  that a given domain can attain. 
The parameter $\delta_{M}$ is an OV concentration threshold, such that  for   $\delta_{k}=\delta_{M}$ the resistivity of the domain $k$ is $\rho_M$. The  sharpness of the transition between $H\rho$ and $L\rho$ is set by the parameter $A$.
We notice that while more involved functional forms could be chosen, Eq.(\ref{eqnre}) nicely captures the behavior  of binary TMOs, where   OV act as n-type dopants supplying free electrons to the conduction band and thereby reducing the electrical resistance.
 We assume that the  parameters $\{\rho_m ,\rho_M, \delta_{M},A \}$ do not change among the different domains, which seems reasonable taking into account that we are modelling  a simple  capacitor-like geometry, with  a single TMO in the MIM structure.
 We distinguish three main regions in the sample (Fig. \ref{Fig1}a): the top and bottom interfaces TI and  BI,  (both defined as interfaces between   the top and bottom electrodes (TE, BE) with the TMO) and the central region C, which represents the bulk zone of the device. In addition,  we define  the active region (AR), which comprises the C and BI. 
   Each  domain $k$ can allocate a concentration of OV, $\delta_k$, with an associated resistivity $\rho_k(\delta_k)$, and  is identified as  a node in the RN which has  horizontal and vertical resistivities,  both equal to $\rho_{k}$. This is equivalent to consider a diagonal resistivity tensor - a well justified assumption in the  absence of magnetic field. 
 Figure \ref{Fig1} b) displays  the dependence of $\rho_{k}$ on $\delta_k$, according to  Eq.(\ref{eqnre}).

Given an external voltage $V$ applied to the TE (the BE is grounded), we   numerically solve
the Kirchhoff´s equations for the RN to obtain
the horizontal and vertical potential drops at each domain $\Delta V_{k,x}$ and $\Delta V_{k,y}$, and compute the total resistance of the  array. 
Following Ref. \onlinecite{roz_2010}, at each simulation step 
the OV profiles are updated according to
{\begin{equation}
    \delta_k (t + \Delta t) = \delta_k (t) +\Sigma_{k'}  \left (p_{(k',k)} - p_{(k,k')}
    \right ) \Delta t,
\end{equation}}
in terms of the probability rates given by
\begin{equation}\label{tr}
    p_{(k,k'),l} = \delta_k (1-\delta_{k'}) e^{-V_\alpha + \Delta V_{k,l}},
\end{equation}
for OV transfer from a domain $k$  to its four nearest neighbours $k'$, with $l= x$, or $l=y$. Note that  the total amount of OV is fixed as  vacancies are not permitted to escape the sample.
The Arrhenius factor contains the local voltage drop $\Delta V_{k,l}$ and the activation energy for OV diffusion $V_\alpha$, both  in units of the  thermal energy $kT$.
The values of $V_\alpha$ are the same along both directions   $x$ and $y$, assuming no specific anisotropies in the sample  that could justify a different choice.
Once the OV concentration at the step ($t + \Delta t$) is calculated,
$\rho_k (t + \Delta t)$ is obtained from Eq.(\ref{eqnre}) and the process starts again.
The model incorporates the presence of  $n_O$ localized defects or static obstacles randomly distributed in the sample. These obstacles  are considered to be places that OV cannot access. This is done by setting the OV concentration at the obstacles to be zero during the whole process. 
As we will discuss below, a key parameter that will determine the dynamics of OV filaments will be the density of obstacles, $\nu_O = \frac{n_O}{N_T}$, where $N_T$ is the total number  of domains.

\section{RESULTS}
\subsection{OV filaments, conductive paths  and resistive changes in the absence of static defects}\label{nodefect}

\begin{figure}[H] 
\centering
\includegraphics[width=1\linewidth]{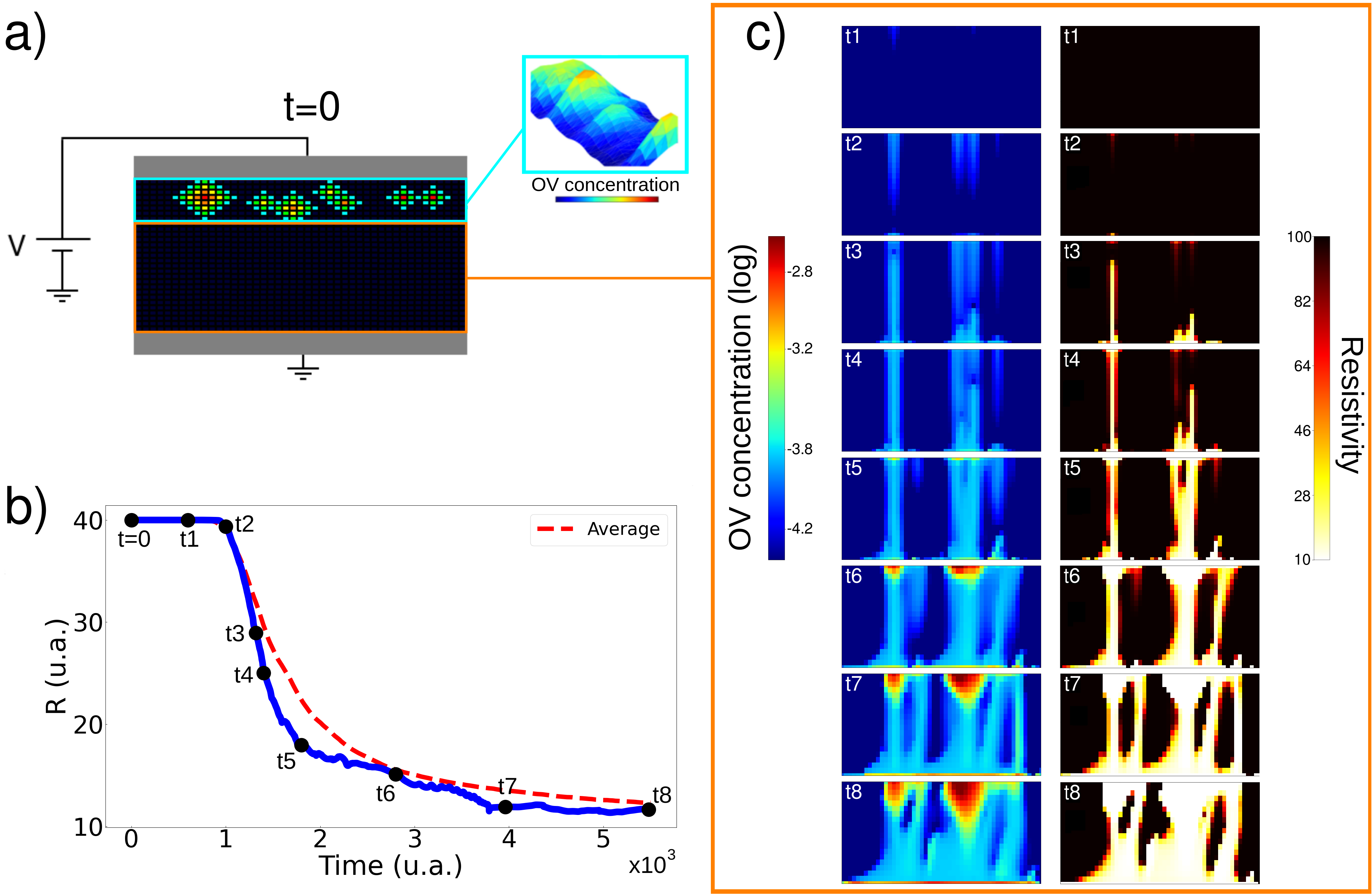}
\caption{(a) Sketch of the  system at $t = 0$. Highlighted by squares are the TI (cyan) and AR (orange). The stimulus $V$ starts at $t = 0$ and is applied to  the TE - the  BE is grounded. The OV concentration´s contour map  and the 3D plot  are also shown. (b) Resistance of the AR as a function of time. The average resistance, taken for a hundred of  realizations is also shown (red dashed line). (c) OV (left) and resistivity (right) profiles along the AR, taken at different times t1 to t8 (shown in (b)).  In all the cases the simulations were ran for  $\delta_M = 1.4 \cdot 10^{-4}$, $A = 10^7$ and  V = 250 a.u. .  }
\label{Fig2}
\end{figure}

In this section we study the dynamics of  OV for a given  positive voltage $V$, in the absence of static defects /obstacles, i.e. $\nu_O = 0$. 
We take an initial OV profile concentrated on the TI, and  obtained from  a randomly generated gaussian-like distribution   - see Fig.\ref{Fig2} a). 
As in the OVRN  we do not consider  anisotropies in the  parameters of the simulations,  the anisotropy in the  initial OV profile along the $x$ direction   acts  as a ``seed" for the formation of OV filaments, as we will discuss in the following. In this respect it can be mentioned that non-uniformities, like grain boundaries, are ubiquitous in amorphous or polycrystalline  oxides- based films \cite{bucki_92}. In addition,  OV profiles in equilibrium often exhibit  a non-uniform distribution along the oxide \cite{lee_2019}.

Figure \ref{Fig2} b) shows the total resistance of the AR  as a function of time (blue line). The  black dots indicate the resistances at  times t1 to t8 (in units of the simulation time) at which the OV  and resistivities profiles  depicted in  Fig.\ref{Fig2} c) were taken.

We define  an OV filament (OVF)  as a path connecting TI and BI, in which each  site has an OV   density $\delta_k \ge 10^{-5}$ . 
This threshold can be slightly modified without affecting qualitatively our results.
In addition   a conductive path (CP) is defined as an OVF in which all the domains are in a  low resistivity state, compared to some threshold (see below). From Eq.(\ref{eqnre})  it is clear  that even after an OVF is formed, if  $\delta_k \ll \delta_M $  for some domains of the filament, the correspondent resistivities  will be  $  \rho_k \sim H_\rho $ and thus the OVF  will not necessarily correspond to a path of low resistance- i.e. a CP. 
To be precise,  to quantify ``low resistivity'' we employ  three different   thresholds, shown in Fig.\ref{Fig1} c). 
The upper one is  the 20$\%$ threshold, meaning  that a  domain with an OV  concentration $\delta_k$ has  a ''low resistivity'' if the condition {$L_\rho\le \rho_k \le H_\rho - 20\% (H_\rho-L_\rho)$ is satisfied. The same reasoning applies to the   $50\%$ and $80\%$ thresholds.}

Different times, t1 to t8, are  chosen in order to characterize the  stages in the formation of the OVFs and the  CPs.  For the smaller times, t1 and t2, the OV concentration  is still  too low and   no  appreciable  change in the resistance  is obtained in Fig.\ref{Fig2} b). Once one (or more than one) OVF is formed, OV start to accumulate at the BI  (see  Fig.\ref{Fig2} c) for  times t3 and t4)  and therefore  one (or more than one) CP  starts to  grow from  the BI to the TI. This is concomitant  with the sharp decrease in the  resistance observed for times t3 to t5. As long as the  external stimulus is  maintained,  new OVFs and CPs can be formed, leading to a further resistance drop that  can be seen  for  times t6 and t7. Eventually, the resistance stabilizes as no new OVF or CP are formed, they only becoming thicker with time (see for example time t8). 

The   formation of  metallic filaments in an environment where metallic ions, like Ag or Ta 
can move easily due to the absence of obstacles  was  also studied  in electrochemical metallization cells (EMC)  \cite{xu_2023, ma_2018, dirkmann_2015, milano_2021, menzel_2015_2}. However, we notice from  our analysis that  an OVF does not necessarily imply  the formation of a CP, a  fact that was disregarded in previous works. 
In our study, the   CP  is formed once the  OV  concentration overcomes a threshold value for all the domains of the OVF and it grows from the BI to the TI, i.e. in    the opposite direction to the OVF growth. 
The present  analysis  can be easily extended to other realizations of the initial OV profile. As an example,  in  Fig.\ref{Fig2} b)  we  plot (red dashed line) the  average   resistance calculated over a hundred of initial (gaussian-like) conditions of the OV profile, showing that it  follows the same  trend as  the  resistance curve corresponding to a single realization (blue curve).
In addition we have checked that besides the random gaussian-like,  other non uniform initial profiles can be employed in order to analyze the formation of OVFs.

\begin{figure}[H] 
\centering
\includegraphics[width=1\linewidth]{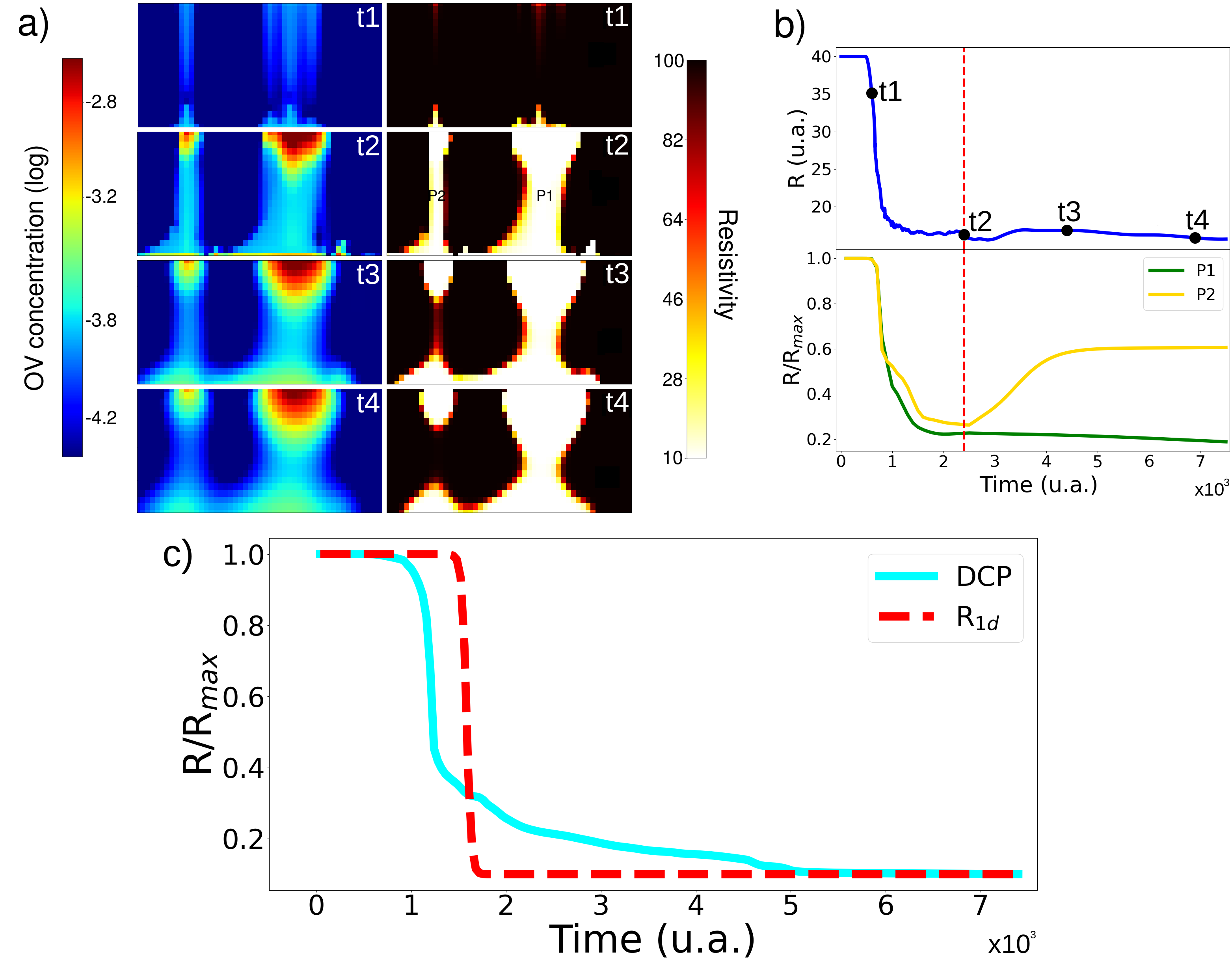}
\caption{(a) Evolution of OV distribution (left) and resistivty (right) profiles. At t1  the electrical stimulus  (V=250 a.u.) is on, and it is switched off at t2. Two CPs, P1 and P2, are identified. (b) Resistance of the AR (top) and of P1 and P2 (bottom), as a function of time. The resistances values at times t1 to t4  are highlighted by the black dots. The dashed vertical line highlight the time t2 when the stimulus is removed. (c) Normalized resistance as a function of time computed numerically for the dominant CP.  Analytic calculation  $R_{1d}$ obtained from Eq.\ref{r1d} (red dashed line). }
\label{Fig3}
\end{figure}

An important issue to consider is the stability  of the OVFs once the external stimulus is removed. Under this scenario, diffusive effects, ruled by the activation  energy $V_\alpha$, start to dominate (see Eq.\ref{tr}) and  could eventually  lead to  the  OVF  dissolution, a phenomenon that is relevant to characterize the volatility of the RS effect \cite{wang_2021, milano_2022, labarbera_2015}. 
In order to analyze the  OVF stability,  we  study  the temporal evolution of the OVFs   under an external  voltage  $V$, which is on during some time and   subsequently removed. We focus on four OV concentration and resistivity profiles displayed in Fig.\ref{Fig3} a) for times ti, (i$=1,4$), indicated in Fig.\ref{Fig3} b).  At time t1 the external stimulus is on, and it is  switch off at time t2,  remaining off  for  the longer times t3 and t4.

As it is shown in Fig.\ref{Fig3} a), and in agreement with the previous  analysis, two OVFs and CPs of different widths are formed between t1 and t2- we denote the thicker CP by P1 and the thinner one by P2, respectively. Once the stimulus is removed, P2 dissolves while P1 remains formed but  acquiring a hourglass-like shape \cite{wang_2021, park_2015}.
The fact that  the  thicker filament  P1 remains formed has a clear effect on the  overall resistance. In Fig.\ref{Fig3} b) we plot the  total resistance  (upper panel) and the   resistances  (properly normalized to their respective maxima) of  both P1 and P2 (bottom panel), as a function of time. Qualitatively, the total resistance follows the same  behaviour as P1, indicating that the thicker CP  dominates the  trend in the overall  resistance. In addition, as we already  showed in Fig.\ref{Fig2} c), OVF and CP widths can be controlled by  the stimulus duration: longer stimulation leads to a lower resistance state, which is  more resilient against diffusive effects \cite{labarbera_2015}.

To further elaborate on  the idea of a dominant CP, in Fig.\ref{Fig3} c)  we plot, for a given realization and for  a constant stimulus V=250 a.u., the resistance of the   OVF  which once stabilized  gives  rise to  the dominant CP. In this case we can approximate the dominant OVF by a  single  1d chain of nanodomains.
Assuming a CP of length L (and section S), and for infinitesimally small domains, one can go to the continuous in the longitudinal dimension and compute  the total resistance employing Eq.(\ref{eqnre}). After replacing  $\delta_i(t) \rightarrow  \delta (x,t)$ for the OV profile at time t,  and neglecting the  geometric  prefactor $1/S$, we get $R_{1d} (t) =  \int_{0}^{L} \rho (x,t) dx = \int_{0}^{L} (\rho_M -\rho_m \tanh[A(\delta(x,t) - \delta_{M})]) dx$ . 


The function $\delta (x,t)$ does not have an analytic expression, so the above integral could only be computed  approximately by assuming a given functional form for the OV profile.
The simplest, rather unjustified assumption, is to consider at each time an uniform profile across the CP:  $\delta(x,t) \equiv \delta_{unif}(t)$. 
 In this case  it is obtained that 
\begin{equation}\label{r1d}
    R_{1d} (t) =L\rho_M - L \rho_m \tanh[A(\delta_{unif} (t) - \delta_{M})] .
\end{equation}
To see how this analytic approximation works, we computed numerically for each  simulation step the total OV concentration along the dominant OVF.  
After dividing it by the total number of sites of the OVF, we  computed   $\delta_{unif} (t)$  and   $R_{1d} (t)$, employing Eq. (\ref{r1d}). The result  is shown by the dashed line in Fig.\ref{Fig3} c).  
Notice that the   behaviour of the  dominant CP resistance  is  quite well captured by $R_{1d} (t)$, particularly  before and after the dominant filament is completely formed.
For intermediate times, when  the  OVF is growing,  the  1d model departs from the dominant CP resistance evolution, as the notion of a ``dominant OVF'' is only well defined once all the filaments have grown. Thus the uniform density approximation misses key information to properly capture the form of $R(t)$ during the transient evolution. The 1d model  also misses out diffusive effects - such as the OVF dissolution shown in Fig. \ref{Fig3}a), and the influence of localized static defects, as we will analyze in the next section.

\begin{figure}[H] 
\centering
\includegraphics[width=1\linewidth]{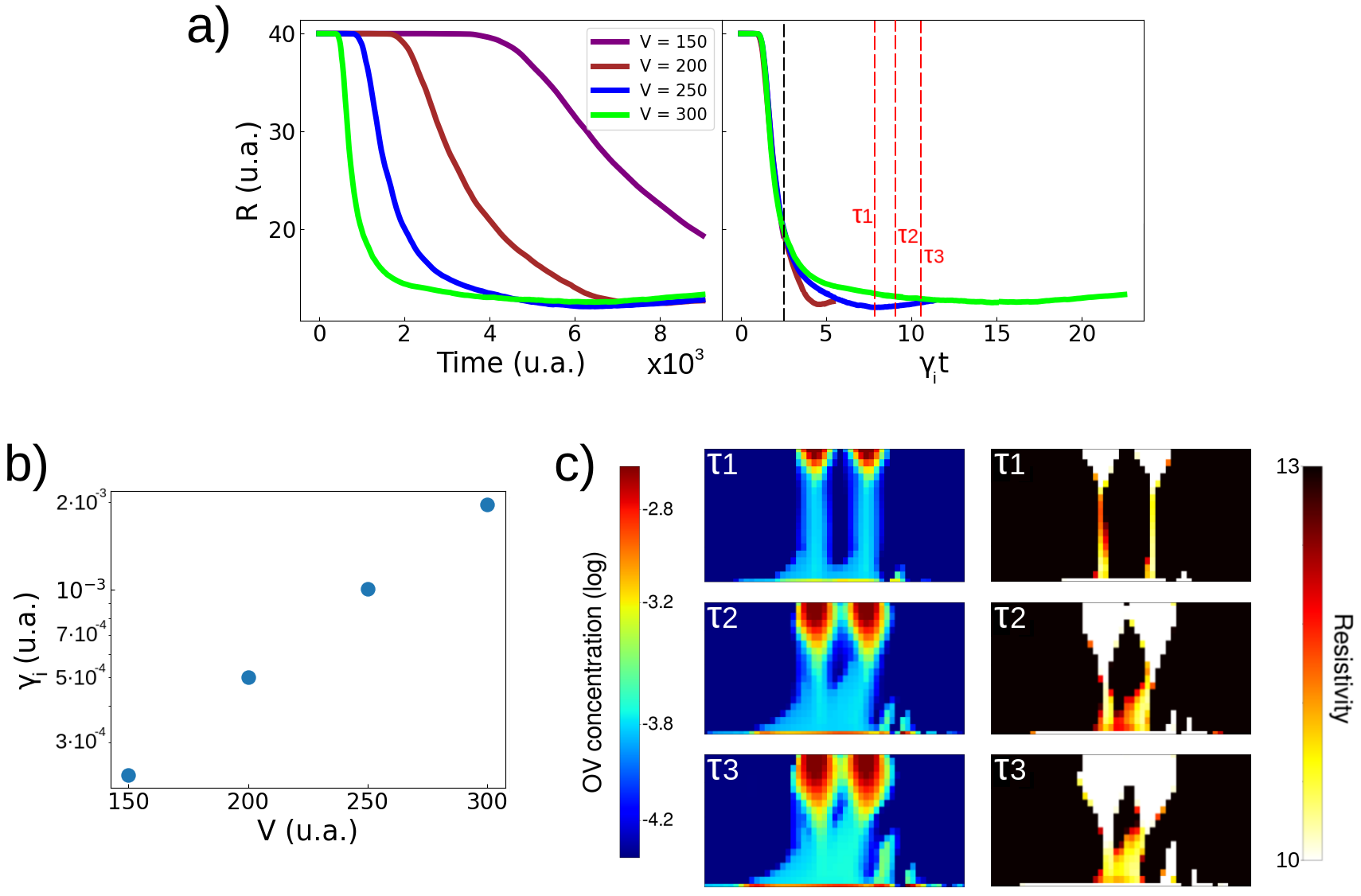}
\caption{a) Evolution of the resistance  for  four different applied voltages $V_i= 150,200, 250, 300$, as a function of time (left) and as a function of $\gamma_i \; t$, $i=1,2,3,4 $ (right). The collapse of the scaled curves is extremely good till  $\gamma_i t\sim 2$ (see the vertical  dashed  black line).  Each curve is an average over nearly 100 realizations of the initial OV profile. b) log-lin plot of $\gamma_i$ as a function of the applied voltages $V_i, i=1,2,3,4$. c) Snapshots of  the OV and resistivity profiles at times $\tau k= \gamma_3 \; t k $,  $k=1,2,3$ ($\gamma_3=1 \cdot 10^{-3}$) indicated in Fig. \ref{Fig4} a) right panel. The profiles are for $V_3$ = 250 a.u and  for a particular  realization of the simulations. All simulations were carried out for $\nu_O = 0$.}
\label{Fig4}
\end{figure}

An interesting outcome of the previous analysis is  the dependence of the  temporal evolution of the resistance with the strength of the applied stimulus. In order to address this issue,  we consider  different applied voltages $V_i$, taking as initial OV concentration the  gaussian-like profiles  previously employed. Figure \ref{Fig4} a) (left panel) shows  the  resistance as a function of time for four different applied voltages (each curve is the average over a hundred realizations of the initial OV profile). In all the cases the initial resistance, $R(t=0)=R_{H}$ is the same. However, the  resistance   drop starts earlier in time  and becomes more pronounced as the value of the voltage increases. In spite of  this behaviour, all the resistance curves collapse onto  a single one after a proper time scaling, $\gamma_i \, t$,
with $\gamma_i\equiv 1/\tau_i$ the inverse of  a characteristic time, $\tau_i$, defined as the  time the resistance remains on its initial (high) value for each voltage $V_i$. 
When plotting  as a function of  $\gamma_i \, t$  all the curves are almost superimposed, as  Fig.\ref{Fig4} a) (right) shows. 
Notice that the collapse is   extremely  good till $\gamma_i \, t \sim 2$ (see the black vertical dashed line in   Fig.\ref{Fig4} a) right panel). This  is  a confirmation of  the  initial exponential decay of  the scaled resistance, as was also  noticed by Tang et al. \cite{tang_2016}.
The scaling factors $\gamma_i$ are shown in Fig. \ref{Fig4} b) for  the four applied voltages (in a log-lin plot),  showing a clear exponential dependence in consistency with the Arrhenius factor appearing in  Eq.(\ref{tr}): the probability rate for  OV migration increases exponentially with the applied voltage, which means that for a  given (constant) $V_i$, the time it takes the OV to reach the AR  to generate the first resistance drop (as shown in Fig.\ref{Fig2}) decreases exponentially  as $V_i$ increases. 
As time further evolves,  the resistance curves reach a minimum due to the 
fact that the OVFs and CPs are completely formed, in consistency with the results  shown in Fig.\ref{Fig2}.
Notice, however, that for longer times- and mainly due to lateral difussion that  tend to  dissolve and eventually merge the already formed OVFs/CPs, a slight increase in the resistance is observed- as can  be   seen in  Fig. \ref{Fig4} a) right panel.  This effect is exemplified   in Fig.\ref{Fig4} c), where  snapshots of the OV and resistivity profiles are shown  for a single realization of the simulations (for $V_3$ = 250 a.u.)  and  for  (dimensionless) times    $\tau k\equiv \gamma_{3} \; t k$,  $k=1,2,3$ defined for $\gamma_{3}$ and  indicated in Fig. \ref{Fig4} a) right panel.
At time $\tau 1$ one identifies  two  completely formed  CPs, in agreement with the  minimum of the  (average) resistance.  However, for longer times diffusion tends to laterally spread  the  OV profiles, slightly increasing the   resistivity values (see the subtle change in the color of the resistivity maps), the effective length of the CPs (see the right CP ''bending'' over to the left) and eventually merging the  CPs. The combination of these effects redounds in an effective (but tiny) increase  of the  (average) resistance  for  $\tau 2$ and $\tau 3$, respectively. 

\begin{figure}[H] 
\centering
\includegraphics[width=1\linewidth]{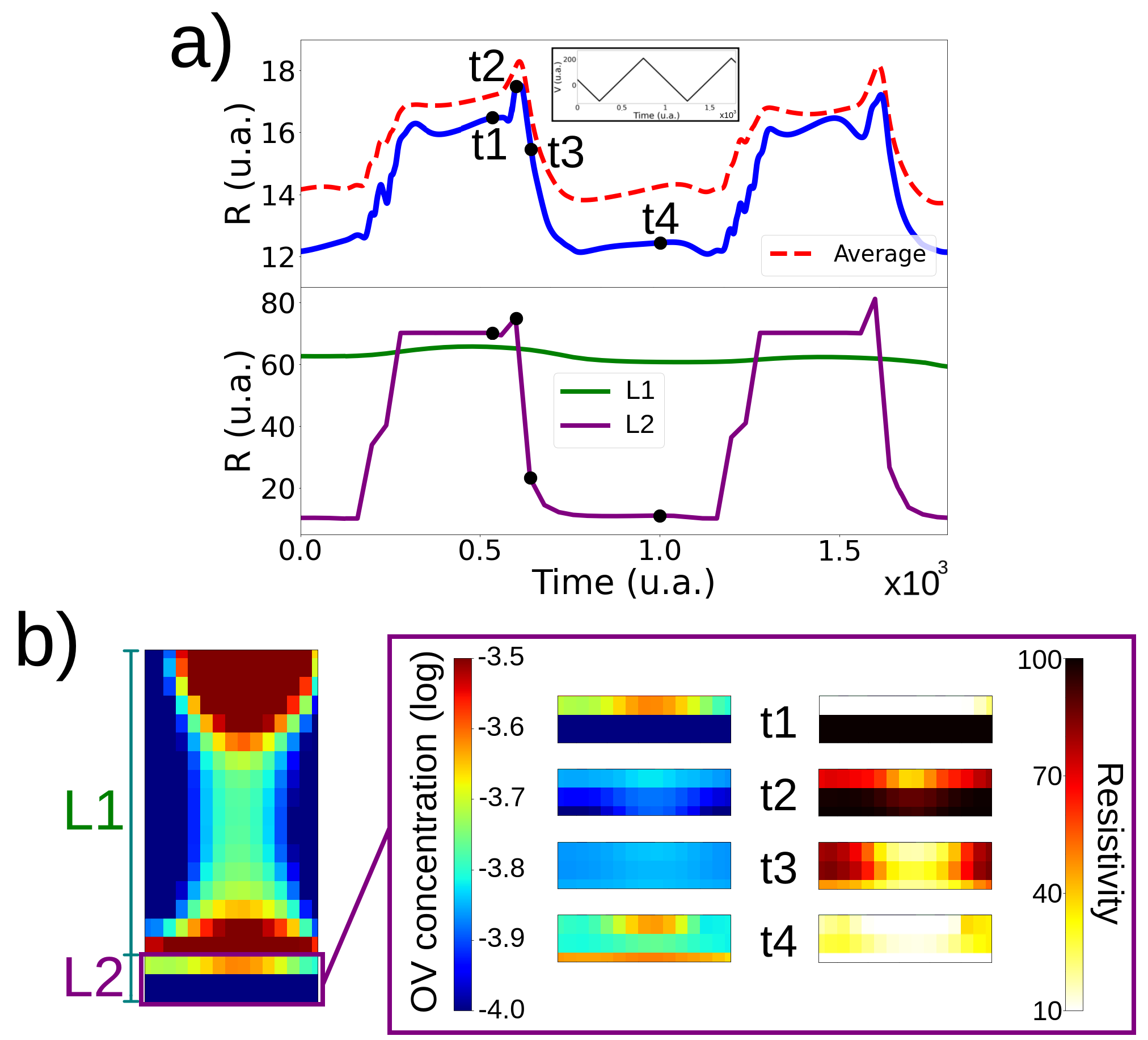}
\caption{a)  Total resistance (top panel)  and  resistances of sections L1 and L2 of the dominant CP (bottom panel), as a function of time. Inset: applied voltage ramp. 
b)  OVF and resistivity profiles  along the section L2 of the   dominant CP for  times t1, t2, t3, t4 indicated in panel a).}
\label{Fig5}
\end{figure}

Inspired by experimental protocols \cite{Fer_2020, park_2015}, we can  account for the RS effect applying a voltage ramp once the OVFs and CPs are formed. 
The results are plotted in  Fig.\ref{Fig5}a)  for  a ramp  with voltage $V$ in the range $-125 \le V\le 205$ a.u.
The  formed OVFs initially  retract due to the  applied negative stimulus and subsequently regrow during   the positive ramp. Along a complete cycle, this retraction/regrowth gives rise to the RS effect, namely, the  commutation of the resistance between  two  values, the high resistance (HR) and the low resistance (LR) state, respectively.
This process is shown in Fig.\ref{Fig5}a) (top panel),  for a single realization of the OV profile (blue line) and for an average over nearly a hundred of realizations (red dashed line). An overall  well defined switching is obtained. In addition, a ``peak'' in the resistance  just before the SET (HR to LR) transition  starts is clearly visible. This peak strongly resembles the  ``abnormal SET''  analyzed by Park et al. in Ref. \onlinecite{park_2015}  and measured in  TaO$_x$ and TiO$_x$-based memristive devices \cite{zhang_2018,park_2017}.

To dive into the origin of this peak, we focus on  the evolution of the dominant CP for the positive ramp ($V\ge 0$) (in the example we plot a single realization). The CP can be separated in two parts, named L1 and L2 (see Fig.\ref{Fig5}b) (left)) and thus the  total resistance  can be  computed as the series resistance of both. In Fig.\ref{Fig5}a) (bottom panel) we show each  resistance  separately,  noticing  that  while L1 resistance stays relatively constant,  L2 resistance  exhibits the peak  observed  for the total resistance in the top panel.
To explain this behaviour we consider  four times, labeled $t1$ to $t4$ (see top panel), and focus on the evolution of  the OV and associated resistivity profiles of L2. The results are shown in Fig.\ref{Fig5}b). At time $t1$, most of the OV are concentrated in the upper part  of L2, and  the resistance is dominated by the bottom part of L2, where no  OV are present  and thus has an associated  high resistivity. Afterwards, the positive applied voltage pushes the  OVs downward and at time $t2$  the OV that were formerly at the top part of L2 are  spilled over the complete  L2 section. In this situation, the resistance of the top part of L2 increases, while the OV density in the bottom part still remains below the threshold $\delta_m$, and no transition to $L\rho$ takes place.

Therefore, this  rearrange of OV on L2   increases the total resistance, and is at the origin of  the peak obtained at  $t2$ before the SET.  After this, OVs  further migrate from L1 to L2 ($t3$ and $t4$) with the concomitant decrease of  the resistance towards  the LR state. It is important to note that, even when OVs migrate from L1 to L2, L1's resistance remains stable due to its OV concentration is above the threshold needed for the transition  to $L\rho$.

In summary the present  analysis,  entirely based in our model for OVF and CP formation, explains in a simple  and direct way the anomalous peak  experimentally detected in the SET transition of (at least) two of the most studied  binary oxide-based memristors with FRS.

\subsection{OVF and CP in the presence of static defects}
\label{sr}
\begin{figure}[H] 
\centering
\includegraphics[width=1\linewidth]{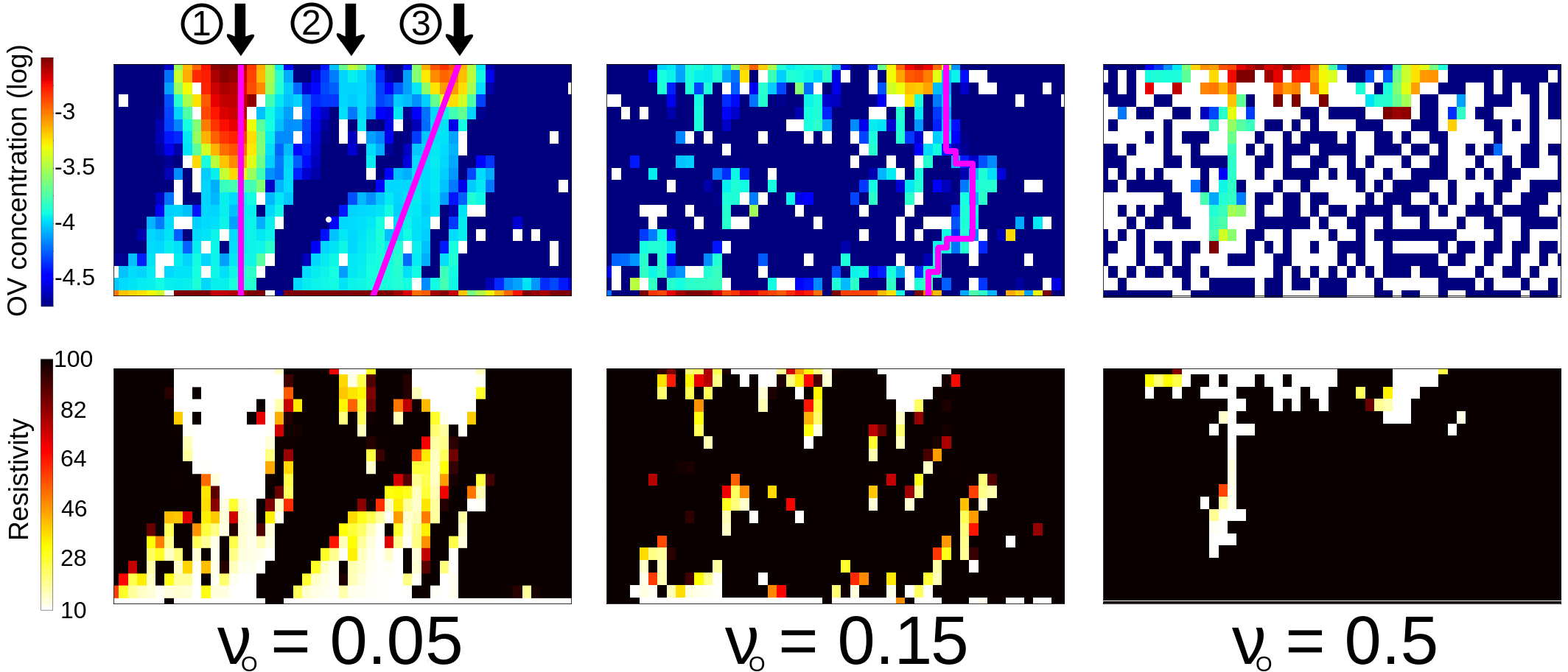}
\caption{OV (top panel) and resistivity (bottom panel) profiles for three  static defect densities $\nu_0=0.05$, $0.15$ and $0.5$, taken once  the profiles became stable in time. The magenta lines are guides to the eye.}
\label{Fig6}
\end{figure}

Adding static defects significantly changes the OVF and CP formation dynamics, as we will discuss in the following.
In Fig.\ref{Fig6} we show  OV
profiles  and the associated  resistivities (bottom panels) for three values of the static defects density $\nu_0 = 0.05$, $0.15$ and $0.5$, respectively. 


For $\nu_0 = 0.05$  we   identify  in the AR three different structures for the OV configuration: The first one corresponds to  OVF and CP  connecting the TI and BI in a  vertical line (the vertical magenta line below the arrow $\textcircled 1$  is shown as a guide to the eye)  similar to what we previously  shown for   $\nu_0 = 0$ in Figs.\ref{Fig2} and \ref{Fig3}.
The second one are OV that remain trapped due to the presence of static defects surrounding them  (see the profile below the  arrow  
$\textcircled 2$ as an example of this) and finally,  "wiggle"  OVF and CP that avoid the randomly distributed static defects (an example of this OV configuration is indicated by the magenta line   plotted as a guide to the eye below   $\textcircled 3$).

Despite these different configurations, we notice that for this small concentration of static defects, an OVF is concomitant with the formation of  a low resistivity CP, as can be seen in the lower panel of Fig.\ref{Fig6} for $\nu_0 = 0.05$.

As the density of localized defects increases,  another regime is obtained. For $\nu_O = 0.15$ for example,  even though a percolating path reaching the BI is formed
 (see  the magenta guiding  line  indicating an OVF formed through such a percolating path), the CP does not  form, as is evident from the  resistivity profiles  displayed in the bottom panel. A sort of  "granular" profile of resistivity is obtained: regions of lower resistivity,  mainly near the TI and the BI, and regions  of higher resistivity, mainly located in the middle of the AR. 
Therefore, an OVF can be formed in this regime, but its OV  density is not  enough to form a CP.

Finally,  the  third regime  corresponds to higher densities $\nu_O$ for which  OV  do not percolate, as exemplified by the $\nu_O = 0.5$ profile. In this regime all OV concentrate near the TI, and  neither an OVF nor  a CP is formed. 

From the precedent discussion it can be inferred  that: i)  the percolation of OV between the TI and BI  is a prerequisite for  the formation of an OVF, and ii) a CP  can be formed  only after  an  OVF is completed. 
Along this reasoning we carried out of the order of a hundred simulations for different static defects concentrations $\nu_O$ to compute the percolation and CP formation probabilities.
The percolation probability is defined in terms of  the fraction of simulations in which there is a path through which OV can percolate from the TI to the BI.
Analogously, the CP formation probability is defined  as the fraction of simulations in which a CP is formed. 
It is  important to note that  we  compute the CP  probability once  the resistivity and the OV profiles have stabilized.

\begin{figure}[H] 
\centering
\includegraphics[width=0.8\linewidth]{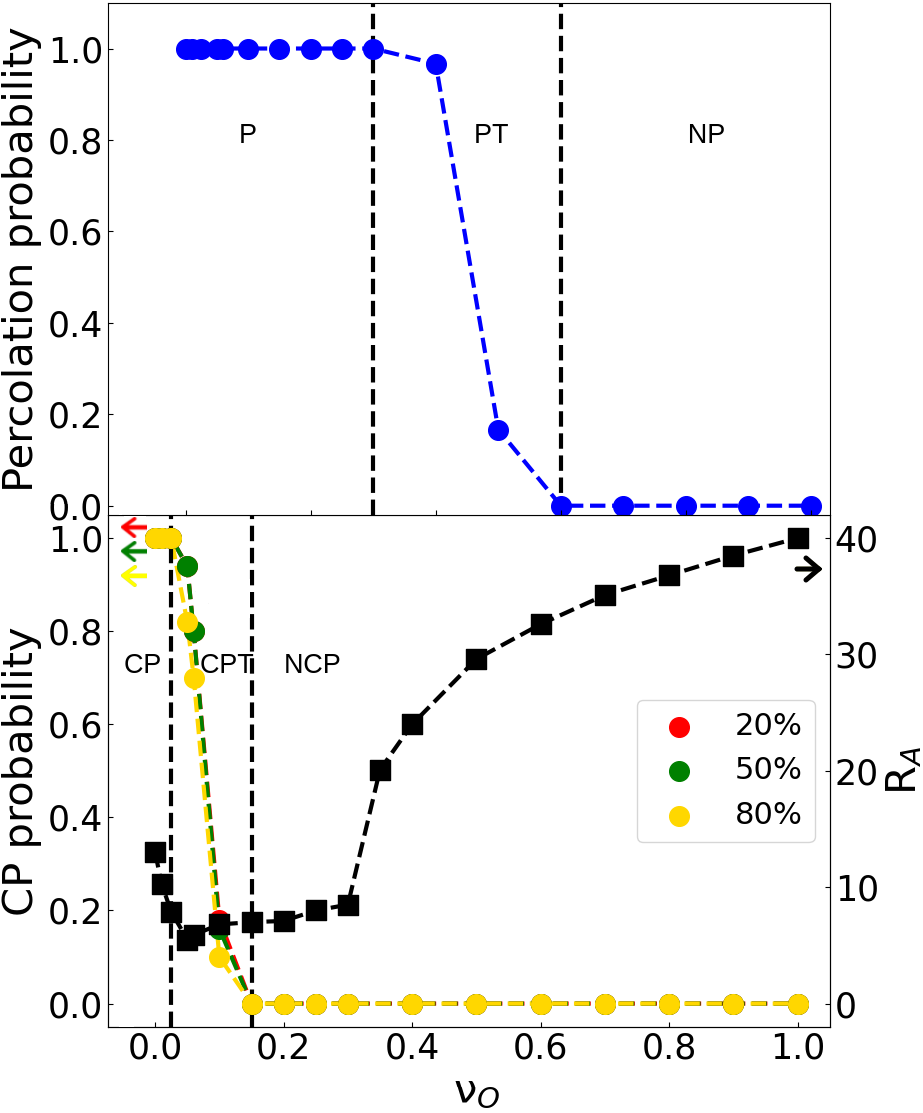}
\caption{Percolation probability (top panel), CP probability (bottom panel, left vertical axis) and  asymptotic resistance $R_{A}$ (bottom panel, right vertical axis) as a function of  static defects concentration $\nu_O$. The CP  probability is plotted for  the three  resisitivity thresholds defined in  Fig. 1 b).  The dashed lines highlight the transition region between the different  regimes described in the text (CP, CPT and NCP, respectively). }

\label{Fig7}
\end{figure}

From our analysis, we identify different thresholds for percolation and CP formation as a function of $\nu_O$.
For $\nu_O\sim 0$,  both the percolating path (P)
and the CP are formed, as can be seen from Fig.\ref{Fig7} (top and bottom panels).
As  $\nu_O$  increases, we obtain  a  regime in which a percolating path is still formed (P)
but  the CP probability decreases. We named this region CPT (see  the bottom panel of the Fig.\ref{Fig7}) and is defined   for values of $\nu_O$  in the interval $(0.025, 0.15)$.  For   larger values,  $0.15 < \nu_O \leq 0.3$,   the percolation probability remains equal to  1 in spite of the fact that the CP probability is negligible (we named this the NCP region). For $0.3 < \nu_O < 0.6$ we define the  PT region, in which the percolation probability decreases 
from 1 to 0. Finally after the PT region, for $\nu_O \geq 0.6$, neither  the OV percolate (NP region) nor  the CP are formed (NCP).

Theoretical works predict, in the thermodynamic limit,  a threshold $p_c= 0.59$  for the onsite percolation transition in square lattices \cite{newman_2000, mertens_2022}. In our model, once we take into account that $\nu_c \equiv 1 - p_c $ we can estimate a value of  $\nu_c \sim 0.45$, which is  in very good agreement  with the theoretical prediction, considering that our system has a finite size.

An important remark is that the present  results are robust and quite independent of the resistivity thresholds employed to define the CP, as can be seen in the three CP probability curves plotted in the bottom panel of Fig.\ref{Fig7}, which  behave following  the same trend. In addition,  since  lowering  $A$ in Eq.(\ref{eqnre})  could lead to  different $\delta_k$ values for each of the selected thresholds, we have checked that  the limits of the CPT region remain roughly unchanged as the value of A is decreased  in one  order of magnitude.

The present analysis highlights  the distinction between a percolating path and a CP. Even when OV percolate through the sample, this  
does not lead necessarily to a CP  connecting  top and bottom electrodes. Indeed, we obtained  a whole range of $\nu_O$ values where OV  percolate without forming a CP but giving rise to the split  profiles in the resistivity maps, as the one   shown in Fig.\ref{Fig6} for $\nu_O=0.15$. In summary, OV percolation is a necessary but not a sufficient condition for the formation of a CP.

\begin{figure}[H] 
\centering
\includegraphics[width=0.85\linewidth]{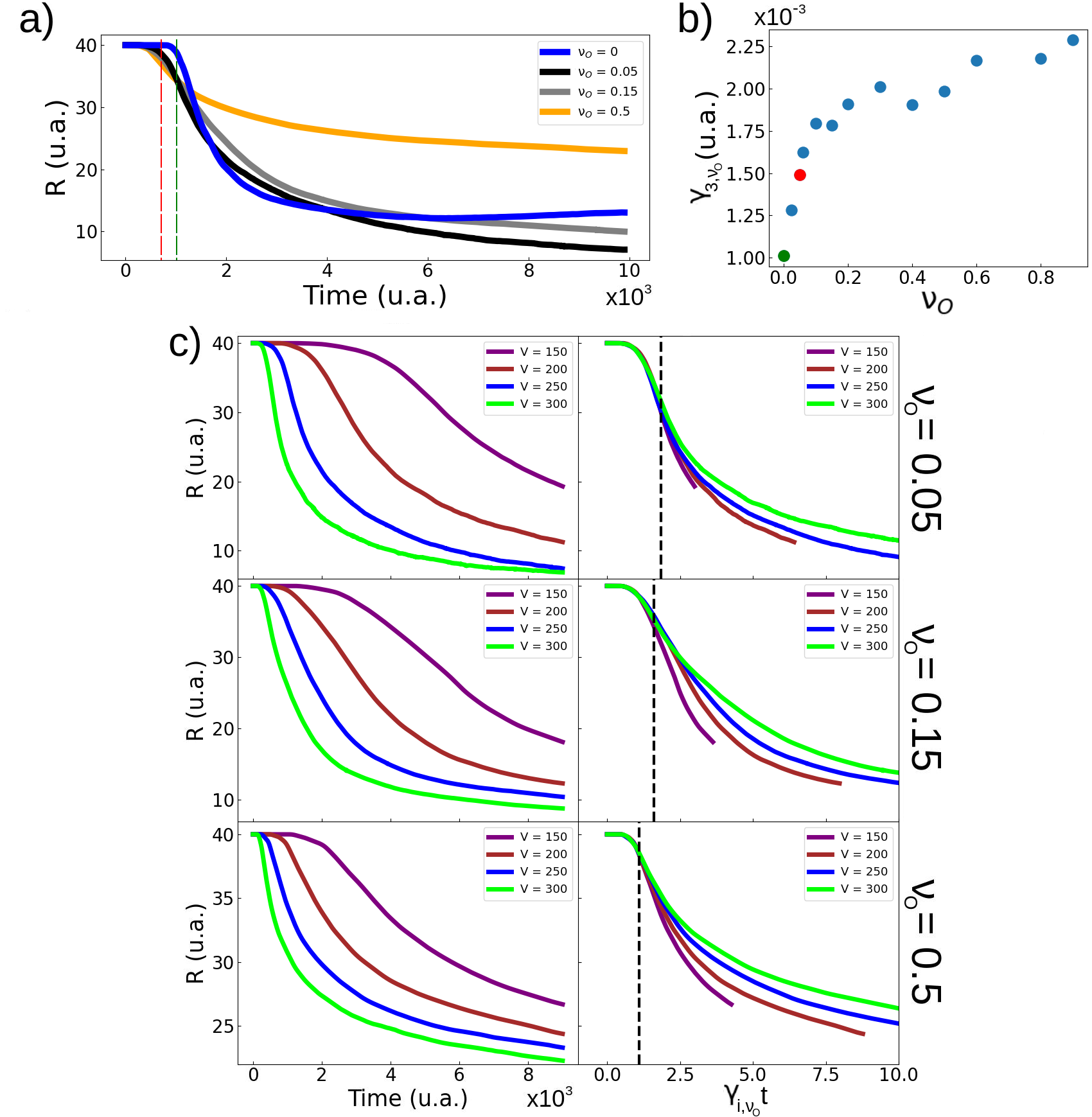}
\caption{a) Resistance as a function of time for four different $\nu_O$ values and for  applied stimulus $V_3$ = 250 a.u. The vertical red and green dashed lines show two different characteristic times  $\tau_{3,0} = 1 \cdot 10^{-3}$ and $\tau_{3,0.05} = 1.6 \cdot 10^{-3}$, corresponding to $\nu_0=0, 0.05$ respectively. b) $\gamma_{3,\nu_O\equiv 1/\tau_{3,\nu_0}}$ values as a function of the static defect density $\nu_O$. The voltage is  $V_3 $= 250 a.u.  c) Resistance as function of time for  $\nu_O= 0.05,0.15,0.5$ (top, central and bottom panel) and  for the four different applied voltages $V_i$ (left panels). The scaled simulations are shown on the right.  In each case the vertical dashed line indicates the scaled time  above which the scaling is lost.} 
\label{Fig8}
\end{figure}

The resistance $R(t)$ changes not only with the applied voltage, as we have already discussed in Sec.\ref{nodefect}, but also with the concentration of static defects $\nu_O$. 
A small concentration of static defects is observed to
reduce the asymptotic  value of the resistance in comparison to the case $\nu_O = 0$ (see for instance in Fig.\ref{Fig8} a) the curve for $\nu_O = 0.05$) -notice however that this occurs without a significant change of the time it takes for $R(t)$ to stabilize . 
This rather paradoxical behavior is systematically observed for the smaller values of $\nu_O$- this  is evident in the asymptotic resistance value, $R_{A}$, computed as a function of the defect concentration in the bottom panel of Fig.\ref{Fig7}, and it originates in the following. A small concentration of defects tends to disturb the movement of vacancies from TI to BI, producing fluctuations in their lateral concentration. These fluctuations promote further migration of vacancies through paths of high OV concentration, additionally reducing the resistance of these paths. In samples without defects there is no such a mechanism to promote the creation of localized low resistance paths. Of course this process can be active only for low enough defect concentration, that do not compromise the global conductivity of the sample.
 As the values of $\nu_O$ further increase along the CPT region,  OVF are still formed but several sites  do not have  the lowest resistivity and as a consequence the overall resistance $R_{A}$ increases with  $\nu_O$. However   the competition  between the average site resistivity and the number of OVFs results in  a smooth increase of  $R_A$.
Finally, for  values of $\nu_O > 0.3$, defects impede the formation of CPs (the  CP probability is negligible) and  a pronounced  increase in the values of $R_{A}$ is obtained.

These different responses  impact on  the temporal evolution of the resistance,   as we discuss in the following.
In Fig.\ref{Fig8} a) we plot  the resistance  as a function of time for several values of $\nu_O$  (the stimulus is $V_3$ = 250 a.u. in all cases).  The first observation is that as $\nu_O$ increases the   resistance drop  starts earlier, as  more defects in the top part of the AR causes OV to accumulate there with the concomitant transition in the lower resistivity values taken place earlier. This effect can be quantified in terms of   the characteristic time  $\tau_{i,\nu_{O}}$ - defined as the time the resistance remains on its initial value for a voltage  $V_i = 150, 200, 250, 300$ ($i = 1,2,3,4$), and static defect density $\nu_{O}$: as $\nu_O$ increases, $\tau_{i,\nu_{O}}$ gets  shorter. The vertical dashed lines in  Fig.\ref{Fig8} a) highlight this result for two characteristic times, $\tau_{3,0}$ and $\tau_{3,0.5}$, respectively. In addition   Fig. \ref{Fig8} b)  shows   $\gamma_{3,\nu_{O}}\equiv 1/\tau_{3,\nu_{O}}$  (for  $V_3$= 250 a.u ($i = 3$))  as a function of $\nu_O$. Despite  the computed  $\gamma_{3,\nu_{O}}$ values exhibit a certain dispersion  (that could be reduced with more statistics),  a clear  (and not linear) growth trend is obtained. 

This analysis has been performed  for  other values of the applied voltages  confirming that static defects  modify the former scaling  with voltage obtained  for the case $\nu_{O}=0$ (and analized in Fig. \ref{Fig4}). 
In Figure\ref{Fig8} c) (left) we show  three vertical panels, each one for a different value $\nu_O=0.05,0.15,0.5 $ respectively, where the resistance is plotted as a function of time for the four different applied stimulus $V_i = 150, 200, 250, 300$ ($i = 1,2,3,4$). On the right the scaled curves are shown. The dashed vertical line in each plot shows the $\gamma_{i,\nu_{O}} t$ at which the scaling is lost:  As more static defects are added into the AR the scaling is lost earlier.




It is worth noting that our $\nu_O$ parameter -related to the amount of blocking nodes present in our 2D resistor network- resembles the porosity $\phi$, defined in a material with voids as $\phi$ = $V_v$ / $V_t$, where $V_v$  and $V_t$ are the voids and total volume, respectively. We recall that the porosity in oxide thin films can reach values up to $\phi$ $\approx$ 0.25 \cite{Bagga_2018}, strongly depending on the fabrication conditions, particularly the growth temperature ($\Phi$ increases as the growth temperature decreases \cite{smith_96}). This shows that, if we assume $\phi$ $\approx$ $\nu_0$,  the range of $\nu_O$ in which we found the crossover from CP to CPT is well within the porosity range reported for oxides \cite{Bagga_2018}, highlighting the potential of our model to help determining the types of nanostructural features compatible with a robust filamentary memristive response.


\section {Concluding remarks}

In this work,  we  analyze  the growth and stability of OV filaments and low resistance conducing paths in  memristive devices based on binary oxides, employing a  2d model which considers the dynamics of OV in  a resistor network. We identify the  main mechanisms involved in these processes noting that, even though  a filament is formed, the OV concentration must exceed a critical value in all the sites in order to generate a low resistivity path. This finding unveils some additional complexity related to filament formation that, to the best of our knowledge, has not been previously addressed.

Our results describe, in addition, a filament formation dynamics that, preceding the transition to a low resistance state, involves the formation of multiple filaments with varying shape and lengths, leading finally to a "dominant filament" -the filament whose resistance follows qualitatively the overall  device resistance. Our model also successfully recreates the RS effect when applying a voltage ramp, giving a simple explanation for an anomalous peak often seen before the SET transition in electrical measurements. The understanding of these features -all correlated with nanoscale OV dynamics- are essential to control the  behavior of filamentary resistive switching devices.

The previous discussion shows that the OVRN model has the potential to further improve our understanding of TMO-based memristive system, noticing that it also helps determining the role of defects on the device electrical response. The inclusion of static defects -corresponding to voids in real (porous) materials- to our simulated system, further highlights the importance of the OVF/CP crossover in the electrical behavior, showing that there is  a whole range of $\nu_O$ values for which OVs percolate through the sample and OVFs are formed, but CPs are absent. This is of paramount importance to optimize the electrical behavior of devices, given the inverse correlation of thin layer porosity and fabrication temperature \cite{Bagga_2018}. Our results allow determining the maximum porosity (correlated with a minimum growth temperature) that warrants the onset of conducting filaments. This is relevant for the development of CMOS-compatible devices, where thermal budget limitations critically constrain the fabrication process. In addition, we also show that the presence of static defects allows the achievement of lower low resistance states,  enlarging in this way the device ON-OFF ratio, which is one of the key figures related to memristive devices design.

We also notice that our model can address the (un)stability of the filaments and conducting paths once the  electrical stimulus is removed (volatile effect), which could be harnessed for the development of neuron-like devices, an issue that  will be addressed in a future work. 

In summary, this paper aims to bridge the gap between theoretical modeling and experimental observations of filamentary resistive switching, offering a two-dimensional perspective that captures the complexities of defect dynamics and conductive filament formation. Our numerical study provides a more comprehensive understanding of filament evolution and its critical role in resistive switching, paving the way for the development and engineering of innovative memristive devices for in-memory or neuromorphic computing that integrate defects as part of their design \cite{baner_2020}.

\textbf{Acknowledgments}
We acknowledge support from ANPCyT (PICT2019-02781, PICT2020A-00415, PICT2019-0564), UNCuyo (P06/C026-T1) and EU-H2020-RISE project MELON (Grant No. 872631).


\textbf{AUTHOR DECLARATIONS}

\textbf{Conflict of Interest}

The authors have no conflicts to disclose.

\textbf{Data availability}

The data that support the findings of this study are available from the corresponding author upon reasonable request.

\bibliography{reference}

\end{document}